\documentstyle[11pt]{article}
\bibliography{plain}
\title{\bf Robustness of entanglement for
two qubit density matrix } \vspace{20mm}
\author{S. J. Akhtarshenas  $^{a,b,c}$
\thanks{E-mail:akhtarshenas@tabrizu.ac.ir} , M. A. Jafarizadeh$^{a,b,c}$
 \thanks{E-mail:jafarizadeh@tabrizu.ac.ir}
\\
\\
$^a${\small Department of Theoretical Physics and Astrophysics,
Tabriz University, Tabriz 51664, Iran.} \\
$^b${\small Institute for Studies in Theoretical Physics and
Mathematics,
Tehran 19395-1795, Iran.} \\
$^c${\small Research Institute for Fundamental Sciences, Tabriz
51664, Iran.}} \pagebreak

\begin{document}
\maketitle \vspace{15mm}
\newpage
\begin{abstract}
By considering the decomposition of a generic two qubit density
matrix presented by Wootters [W. K. Wootters, Phys. Rev. Lett.
{\bf 80} 2245 (1998)], the robustness of entanglement for any
mixed state of two qubit systems is obtained algebraically. It is
shown that the robustness of entanglement is proportional to
concurrence and in Bell decomposable density matrices it is equal
to the concurrence. We also give  an analytic expression for two
separable states which wipe out all entanglement of these states.
Since thus obtained robustness is function of the norm of the
vectors in the decomposition we give an explicit parameterization
for the decomposition.


{\bf PACs Index: 03.65.Ud }

\end{abstract}
\pagebreak

\vspace{7cm}

\section{Introduction}
Quantum entanglement has recently been attracted much attention as
a potential resource for communication and information processing
\cite{ben1,ben2}. Entanglement is usually arise from quantum
correlations between separated subsystems which can not be created
by local actions on each subsystem. By definition, a mixed state
$\rho$ of a bipartite system is said to be separable (non
entangled) if it can be written as a convex combination of product
states
$$
\rho=\sum_{i}w_{i}\,\rho_i^{(1)}\otimes\rho_i^{(2)},\qquad w_i\geq
0, \quad \sum_{i}w_i=1,
$$
where $\rho_i^{(1)}$ and $\rho_i^{(2)}$ are states of subsystems
$1$ and $2$, respectively. Although, in the case of pure states of
bipartite systems it is easy to check whether a given state is, or
is not entangled, the question is yet an open problem in the case
of mixed states. Therefore having a measure to quantify
entanglement of mixed states is likely to be valuable and several
measures of entanglement have been proposed
\cite{ben3,ved1,ved2,woot}.

One useful quantity introduced  in \cite{vidal} as a measure of
entanglement is robustness of entanglement. It corresponds to the
minimal amount of mixing with locally prepared states which washes
out all entanglement. Authors in \cite{vidal} have given
analytical expression for pure states of binary systems. A
geometrical interpretation of robustness is given in \cite{du} and
have pointed that two corresponding separable states needed to
wipe out all entanglement are necessarily on the boundary of
separable set. Unfortunately, the above mentioned quantity as the
most proposed measures of entanglement involves exteremization
which are difficult to handel analytically.

In this paper we evaluate robustness for a generic two qubit
density matrix. Our approach to finding robustness is based on the
decomposition given by Wootters in \cite{woot}. Wootters in
\cite{woot} has shown that for any two qubit density matrix there
always exist the decomposition
$\rho=\sum_i\left|x_i\right>\left<x_i\right|$ in such a way that
$\left<x_i|\tilde{x}_j\right>=\lambda_i\delta_{ij}$ where
$\lambda_i$ are square roots of eigenvalues, in decreasing order,
of the non-Hermitian matrix $\rho\tilde{\rho}$. We consider
Wootters decomposition and show that for a generic two qubit
density matrix we can associate a tetrahedral with coordinates
$P_i=\left<x_i|x_i\right>$. It is shown that in terms of these
coordinates the corresponding separable states form an irregular
octahedral. Based on this we evaluate algebraically robustness of
entanglement of a generic two qubit density matrix by using a new
norm defined by Wootters's basis. We also give an analytic
expression for separable states that wipe out all entanglement and
show that they are on the boundary of separable states as pointed
out in \cite{du}. Since in our approach of evaluation of
robustness, one needs to know the norm of the vectors in the
decomposition, we give an explicit parameterization for the
Wootters decomposition to evaluate the norm explicitly.

The paper is organized as follows. In section 2 using the
Wootters's basis \cite{woot} we define a new norm to evaluate
robustness. Robustness of entanglement of a generic two qubit
density matrix is evaluated in section 3. By giving an explicit
parameterization for density matrix $\rho$, the norm of the
vectors in Wootters decomposition of $\rho$ is evaluated in
section 4. The paper is ended with a brief conclusion in section
5.

\section{Physical norm based on Wootters's basis}
Here in this section we define a new norm to evaluate robustness
of a generic two qubit density matrix. To this aim we first review
concurrence and Wootters's basis as presented by Wootters in
\cite{woot}. Wootters in \cite{woot} has shown that for any two
qubit density matrix $\rho$ there always exist a decomposition
\begin{equation}\label{rhox}
\rho=\sum_i\left|x_i\right>\left<x_i\right|,
\end{equation}
called Wootters's basis, such that
\begin{equation}\label{xxt}
\left<x_i|\tilde{x}_j\right>=\lambda_i\delta_{ij},
\end{equation}
where $\lambda_i$ are square roots of eigenvalues, in decreasing
order, of the non-Hermitian matrix $\rho\tilde{\rho}$ and
\begin{equation}\label{rhotilde}
{\tilde \rho}
=(\sigma_y\otimes\sigma_y)\rho^{\ast}(\sigma_y\otimes\sigma_y),
\end{equation}

where $\rho^{\ast}$ is the complex conjugate of $\rho$ when it is
expressed in a standard basis such as
$\{\left|\uparrow\uparrow\right>,
\left|\uparrow\downarrow\right>\},\{\left|\downarrow\uparrow\right>,
\left|\downarrow\downarrow\right>\}$ and $\sigma_y$ represent
Pauli matrix in local basis $\{\left|\uparrow\right>,
\left|\downarrow\right>\}$ . Based on this, the concurrence of the
mixed state $\rho$ is defined by
$\max(0,\lambda_1-\lambda_2-\lambda_3-\lambda_4)$ \cite{woot}.

Now let us define states $\left|x^\prime_i \right>$ as
\begin{equation}\label{xprime}
\left|x^\prime_i \right>=\frac{\left|x_i
\right>}{\sqrt{\lambda_i}},\qquad \mbox{for}\,\,i=1,2,3,4.
\end{equation}
Then $\rho$ can be expanded as
\begin{equation}\label{rholambda}
\rho=\sum_i\lambda_i\left|x^\prime_i \right>\left<x^\prime_i
\right|,
\end{equation}
and Eq. (\ref{xxt}) takes the following form
\begin{equation}\label{xpxpt}
\left<x^\prime_i|\tilde{x^\prime_j}\right>=\delta_{ij}.
\end{equation}
Using the independency of Wootters's basis, one can expand an
arbitrary $4\times 4$ Hermitian matrix $M$ in terms of them, that
is, we can write
\begin{equation}\label{Mxpxp}
M=\sum_{ij}a_{ij}|x^\prime_i\left>\right<x^\prime_j|,
\end{equation}
where the Hermiticity  of $M$ implies that $a_{ij}=a^\ast_{ji}$.
We now define norm of the matrix $M$ by
\begin{equation}\label{newnorm}
\|M\|=\sqrt{|Tr(M\tilde{M})|}=\sqrt{|\sum_{ij}a_{ij}^2|},
\end{equation}
where $\tilde{M}$ is defined as Eq. (\ref{rhotilde}). Obviously,
the absolute value appearing in Eq. (\ref{newnorm}) guarantees the
positivity of the above defined norm. In fact in cases that $M$ is
nonnegative matrix then ${\tilde M}$ and $M{\tilde M}$ is also
nonnegative and absolute value sign in Eq. (\ref{newnorm}) can be
neglected.\\ Any Physical norm in quantum information should be at
least invariant under local unitary transformation and below we
prove that the norm defined in Eq. (\ref{newnorm}) is actually
invariant under local unitary transformation. Suppose the matrix
$M$ is subjected to local unitary transformation defined as
\begin{equation}\label{MMp}
M\longrightarrow M^\prime=(U_1\otimes U_2)M(U_1\otimes U_2)^\dag
=\sum_{ij}a_{ij}|x^{\prime\prime}_i\left>\right<x^{\prime\prime}_j|,
\end{equation}
where $\left|x^{\prime\prime}_i\right>=(U_1\otimes U_2)
\left|x^{\prime}_i\right>$ are the new Wootters's basis with the
corresponding $\left|\tilde{x^{\prime\prime}_i}\right>$ defined as
\begin{equation}\label{xppt}
\left|\tilde{x^{\prime\prime}_i}\right>=(\sigma_y\otimes\sigma_y)(U^\ast_1\otimes
U^\ast_2)(\sigma_y\otimes\sigma_y)\left|\tilde{x^{\prime}_i}\right>
\end{equation}
which satisfy
\begin{equation}\label{xppxppt}
\left<x^{\prime\prime}_i|\tilde{x^{\prime\prime}_j}\right>=\delta_{ij},
\end{equation}
where in Eq. (\ref{xppxppt}) we have used the fact that
$(U^T\sigma_y U)_{ij}
=-i\epsilon_{kl}U_{ki}U_{lj}=-i\det(U)\epsilon_{ij}=(\sigma_y)_{ij}$.
Therefore using the definition of norm given in Eq.
(\ref{newnorm}) and using Eq. (\ref{xppxppt}) we have
\begin{equation}\label{newnormLU}
\|M^\prime\|=\sqrt{|\sum_{ij}a_{ij}^2|}
\end{equation}
Now, comparing Eq. (\ref{newnormLU}) with (\ref{newnorm}) we see
that norm of $M$ is invariant under local unitary
transformations.\\
 With these considerations it is natural to define
distance between two density matrices $\rho_1$ and $\rho_2$ as
\begin{equation}\label{newdis}
\|\rho_1-\rho_2\|=\sqrt{\left|Tr((\rho_1-\rho_2)({\tilde
\rho_1}-{\tilde \rho_2}))\right|}
\end{equation}
which is invariant under local unitary transformations
.
\section{Robustness for two qubit density matrices}
According to \cite{vidal} for a given entangled state $\rho$ and
 separable state $\rho_{s}$, a new density matrix $\rho(s)$
 can be constructed as,
 \begin{equation}\label{rhos}
 \rho(s)=\frac{1}{s+1}(\rho+s\rho_s),\quad s\geq0,
 \end{equation}
 where it can be either entangled or separable.
 It was pointed that there always exits the minimal ${\bf s}$
 corresponding to $\rho_s$ such that $\rho(s)$ is separable. This
 minimal ${\bf s}$ is called the robustness of $\rho$ relative to
 $\rho_s$, denoted by $R(\rho\parallel\rho_s)$. The absolute
 robustness of $\rho$ is defined as the quantity,
 \begin{equation}\label{rob1}
 R(\rho\parallel S)\equiv\min_{\rho_s\in S} R(\rho\parallel
 \rho_s).
  \end {equation}

Du et al. in \cite{du} have given a geometrical interpretation of
robustness and pointed that if ${\bf s}$ in Eq. (\ref{rhos}) is
minimal among all separable states $\rho_s$, i.e. ${\bf s}$ is the
absolute robustness of $\rho$, then $\rho_s$ and $\rho(s)$ in Eq.
(\ref{rhos}) are necessarily on the boundary of the separable
states.

Here in this section we obtain robustness for a generic two qubit
density matrix. Our method of evaluation of robustness is based on
the decomposition of density matrix given by Wootters in
\cite{woot}. By defining $P_i=\lambda_iK_i$, where
$K_i=\left<x^\prime_i|x^\prime_i\right>$, then normalization
condition of $\rho$ leads to
\begin{equation}\label{normal}
Tr(\rho)=\sum_{i=1}^{4}P_i=1, \qquad P_i>0.
\end{equation}
This means that with respect to coordinates $P_i$, the space of
density matrices forms a tetrahedral. With respect to this
representation separability condition
$\lambda_1-\lambda_2-\lambda_3-\lambda_4\leq 0$ takes the
following form
\begin{equation}\label{sepcon}
\frac{P_1}{K_1}-\frac{P_2}{K_2} -\frac{P_3}{K_3}
-\frac{P_4}{K_4}\leq 0.
\end{equation}
States that saturate inequality (\ref{sepcon}) form a plane called
${\cal S}_1$ (see Fig. 1). All states violating inequality
(\ref{sepcon}) are entangled states for which $\lambda_1\geq
\lambda_2+\lambda_3+\lambda_4$. These states form an entangled
region with ${\cal S}_1$ as its separable boundary. There exist,
however, three other entangled regions corresponding with the
cases that $\lambda_j$ is dominated ($j=2,3,4$). These regions
also define separable planes ${\cal S}_j$. Four planes ${\cal
S}_i$ together with four planes ${\cal S}^{\prime}_i$,
corresponding to $\lambda_i=0$, form an irregular octahedral
corresponding to the separable states. This geometry is similar to
that of Bell decomposable states but here we have an irregular
octahedral associated to separable states \cite{horo}. Figure 1
shows a perspective of this geometry, where two separable planes
${\cal S}_1$ and ${\cal S}^{\prime}_1$ are shown explicitly.

Now in order to obtain robustness of $\rho$ suppose that a ray
from $\rho$ is drawn such that intersects the boundary planes of
separable region at points $\rho^\prime_s$ and
$\rho^{\prime\prime}_s$. Although $\rho^\prime_s$ is necessarily
on the plane ${\cal S}_1$, but $\rho^{\prime\prime}_s$ is allowed
to lie on any plane ${\cal S}^{\prime}_1$, ${\cal S}_2$, ${\cal
S}_3$ or ${\cal S}_4$, where we evaluate robustness for each case
separately. First consider the case that $\rho^{\prime\prime}_s$
is lying on the plane ${\cal S}^{\prime}_1$. In this case
$\rho^{\prime\prime}_s$ can be written as a convex sum of three
vertices of the plane
\begin{equation}\label{sigma234}
\rho^{\prime\prime}_s
=\sum_i\lambda^{\prime\prime}_i\left|x^\prime_i\right>\left<x^\prime_i\right|
=a_2\sigma_2+a_3\sigma_3+a_4\sigma_4,
\qquad a_2+a_3+a_4=1,
\end{equation}
where $\sigma_i$ are separable states that can be written as
convex sum of two corresponding vertices of tetrahedral as
\begin{equation}\label{sigma2}
\sigma_2=\frac{1}{K_3+K_4}\left|x^{\prime}_3\right>\left<x^{\prime}_3\right|
+\frac{1}{K_3+K_4}\left|x^{\prime}_4\right>\left<x^{\prime}_4\right|,
\end{equation}
\begin{equation}\label{sigma3}
\sigma_3=\frac{1}{K_2+K_4}\left|x^{\prime}_2\right>\left<x^{\prime}_2\right|
+\frac{1}{K_2+K_4}\left|x^{\prime}_4\right>\left<x^{\prime}_4\right|,
\end{equation}
\begin{equation}\label{sigma4}
\sigma_4=
\frac{1}{K_2+K_3}\left|x^{\prime}_2\right>\left<x^{\prime}_2\right|
+\frac{1}{K_2+K_3}\left|x^{\prime}_3\right>\left<x^{\prime}_3\right|,
\end{equation}
and $\lambda^{\prime\prime}_i$ are
\begin{equation}\label{lambda1pp}
\lambda^{\prime\prime}_1=0,
\end{equation}
\begin{equation}\label{lambda2pp}
\lambda^{\prime\prime}_2=\frac{a_3}{K_2+K_4}+\frac{a_4}{K_2+K_3},
\end{equation}
\begin{equation}\label{lambda3pp}
\lambda^{\prime\prime}_3=\frac{a_2}{K_3+K_4}+\frac{a_4}{K_2+K_3},
\end{equation}
\begin{equation}\label{lambda4pp}
\lambda^{\prime\prime}_4=\frac{a_2}{K_3+K_4}+\frac{a_3}{K_2+K_4}.
\end{equation}
By expanding $\rho^{\prime}_s$ as convex sum of $\rho$ and
$\rho^{\prime\prime}_s$
\begin{equation}\label{convexsum}
\rho^{\prime}_s=\frac{1}{1+s}(\rho+s\rho^{\prime\prime}_s),
\end{equation}
and also using the fact that the coordinates of $\rho^{\prime}_s$
satisfy the equation
\begin{equation}\label{rhopscond}
\frac{P^{\prime}_1}{K_1}-\frac{P^{\prime}_2}{K_2}-\frac{P^{\prime}_3}{K_3}
-\frac{P^{\prime}_4}{K_4}=0,
\end{equation}
after some algebra coordinates $\lambda^\prime_i$ of
$\rho^\prime_s=\sum_i\lambda^{\prime}_i\left|x^\prime_i\right>\left<x^\prime_i\right|$
can be written as
\begin{equation}\label{lambda1p}
\lambda^\prime_1=
\frac{\left(\frac{a_2}{K_3+K_4}+\frac{a_3}{K_2+K_4}+\frac{a_4}{K_2+K_3}\right)\lambda_1}
{\frac{a_2}{K_3+K_4}+\frac{a_3}{K_2+K_4}+\frac{a_4}{K_2+K_3}+\frac{C}{2}},
\end{equation}
\begin{equation}\label{lambda2p}
\lambda^\prime_2=
\frac{\left(\frac{a_2}{K_3+K_4}+\frac{a_3}{K_2+K_4}+\frac{a_4}{K_2+K_3}\right)\lambda_2+
\frac{C}{2}\left(\frac{a_3}{K_2+K_4}+\frac{a_4}{K_2+K_3}\right)}
{\frac{a_2}{K_3+K_4}+\frac{a_3}{K_2+K_4}+\frac{a_4}{K_2+K_3}+\frac{C}{2}},
\end{equation}
\begin{equation}\label{lambda3p}
\lambda^\prime_3=
\frac{\left(\frac{a_2}{K_3+K_4}+\frac{a_3}{K_2+K_4}+\frac{a_4}{K_2+K_3}\right)\lambda_3+
\frac{C}{2}\left(\frac{a_2}{K_3+K_4}+\frac{a_4}{K_2+K_3}\right)}
{\frac{a_2}{K_3+K_4}+\frac{a_3}{K_2+K_4}+\frac{a_4}{K_2+K_3}+\frac{C}{2}},
\end{equation}
\begin{equation}\label{lambda4p}
\lambda^\prime_4=
\frac{\left(\frac{a_2}{K_3+K_4}+\frac{a_3}{K_2+K_4}+\frac{a_4}{K_2+K_3}\right)\lambda_4+
\frac{C}{2}\left(\frac{a_2}{K_3+K_4}+\frac{a_3}{K_2+K_4}\right)}
{\frac{a_2}{K_3+K_4}+\frac{a_3}{K_2+K_4}+\frac{a_4}{K_2+K_3}+\frac{C}{2}},
\end{equation}
where $C=\lambda_1-\lambda_2-\lambda_3-\lambda_4$ is the
concurrence of $\rho$. By using the above result and the
definition of distance given in Eq. (\ref{newdis}) one can
evaluate robustness of $\rho$ relative to $\rho^{\prime\prime}_s$
as
\begin{equation}\label{s1}
s_1=\frac{\|\rho-\rho^\prime_s\|}{\|\rho^\prime_s-\rho^{\prime\prime}_s\|}=
\sqrt{\frac{\sum_i(\lambda_i-\lambda^\prime_i)^2}
{\sum_i(\lambda^\prime_i-\lambda^{\prime\prime}_i)^2}}
=\frac{C}{\frac{2a_2}{K_3+K_4}+\frac{2a_3}{K_2+K_4}+\frac{2a_4}{K_2+K_3}}.
\end{equation}
Analogue to the above method one can evaluate robustness of $\rho$
for the case that $\rho^{\prime\prime}_s$ lies on the plane ${\cal
S}_2$. In this case $\rho^{\prime\prime}_s$ can be expanded as
convex sum of three vertices of the plane
\begin{equation}\label{sigma123}
\rho^{\prime\prime}_s=b_1\sigma_1+b_3\sigma_3+b_4\sigma_4, \qquad
b_1+b_3+b_4=1,
\end{equation}
where
\begin{equation}\label{sigma1}
\sigma_1=\frac{1}{K_1+K_2}\left|x^{\prime}_1\right>\left<x^{\prime}_1\right|
+\frac{1}{K_1+K_2}\left|x^{\prime}_2\right>\left<x^{\prime}_2\right|,
\end{equation}
and $\sigma_3$ and $\sigma_4$ are defined in Eqs. (\ref{sigma3})
and (\ref{sigma4}). Then after some algebra we obtain the
corresponding robustness as
\begin{equation}\label{s2}
s_2=\frac{C}{\frac{2b_3}{K_2+K_4}+\frac{2b_4}{K_2+K_3}}.
\end{equation}
Similarly in cases that separable state $\rho^{\prime\prime}_s$
are on the planes ${\cal S}_3$ and ${\cal S}_4$ we obtain relative
robustness of $\rho$ as
\begin{equation}\label{s3}
s_3=\frac{C}{\frac{2c_2}{K_3+K_4}+\frac{2c_3}{K_2+K_4}},
\end{equation}
and
\begin{equation}\label{s4}
s_4=\frac{C}{\frac{2d_3}{K_2+K_4}+\frac{2d_4}{K_2+K_3}},
\end{equation}
respectively. Equations. (\ref{s1}), (\ref{s2}), (\ref{s3}) and
(\ref{s4}) show that in order to achieve  the minimum robustness
it is enough to consider the case that separable state
$\rho^{\prime\prime}_s$ lies on the plane ${\cal S}^{\prime}_1$.
With this consideration we are now allowed to choose coefficients
$a_i$ in such a way that Eq. (\ref{s1}) becomes minimum. It is
easy to see that this happens as long as the coefficient $a_k$
corresponding to the term $\min(K_i+K_j)$ becomes one. Therefore
robustness of $\rho$ relative to $\rho^{\prime\prime}_s$ is
\begin{equation}\label{rob2}
s=\frac{\min(K_i+K_j)}{2}C
\end{equation}
which is the main result this work. Here the  minimum is taken
over all combination of $K_i+K_j$ for $i,j=2,3,4$. Equation
(\ref{rob2}) implies that for two qubit systems robustness is
proportional to the concurrence. We see that the minimum
robustness given in Eq. (2-3) corresponds to $a_i=\delta_{ik}$,
therefore, by using Eq. (\ref{sigma234}) we get the following
result for $\rho^{\prime\prime}_s$
\begin{equation}\label{rhoppsigma}
\rho^{\prime\prime}_s=\sigma_k.
\end{equation}
Also by using $a_i=\delta_{ik}$ in Eqs. (\ref{lambda1p}) to
(\ref{lambda4p}) one can easily obtain the coordinates
$\lambda^\prime_i$ of separable state $\rho^{\prime}_s$.

As we will show in the next section the Bell decomposable states
correspond to the $K_i=1$ for $i=1,2,3,4$, therefore in Bell
decomposable states Eq. (\ref{rob2}) implies that the robustness
is equal to the concurrence.

Now we have to show that thus obtained robustness is minimum over
all separable states. Vidal et al. in \cite{vidal} have shown that
$R(\rho\parallel\rho_s)$ is a convex function of $\rho_s$. This
means that any local minimum is also the absolute one, thus in
order to find the absolute minimum of $R(\rho\parallel\rho_s)$ as
a function of $\rho_s$ it is enough to find local minimum
\cite{vidal}. As the above mentioned argument shows robustness
given in Eq. (\ref{rob2}) is minimal relative to all separable
states of the tetrahedral. In the rest of this section we want to
show that it is indeed minimum over all separable states. To this
aim let us consider following pseudomixture for the density matrix
$\rho$ given in Eq. (\ref{rholambda})
\begin{equation}\label{pseudoAB}
\rho=(1+s)\rho^\prime_s-s\,\rho^{\prime\prime}_s,
\end{equation}
where $\rho^\prime_s$ and $\rho^{\prime\prime}_s$ are two
separable states with following decomposition
\begin{equation}\label{rhopsA}
\rho^\prime_s=\sum_{i}\lambda^\prime_i\left|x^\prime_i\left>\right<x^\prime_i\right|
+\sum_{i,j}a_{ij}|x^\prime_i\left>\right<x^\prime_j|,
\end{equation}
\begin{equation}\label{rhoppsB}
\rho^{\prime\prime}_s=
\sum_i\lambda^{\prime\prime}_i\left|x^\prime_i\right>\left<x^\prime_i\right|
+\sum_{i,j}b_{ij}|x^\prime_i\left>\right<x^\prime_j|,
\end{equation}
where $A$ and $B$ are, respectively, off-diagonal Hermitian
matrices $A=(a_{ij})$ and $B=(b_{ij})$. It follows from Eqs.
(\ref{pseudoAB}) to (\ref{rhoppsB}) that the following equations
should hold
\begin{equation}\label{eq1AB}
\lambda_i=(1+s)\lambda^{\prime}_i-s\,\lambda^{\prime\prime}_i,
\end{equation}
\begin{equation}\label{eq2AB}
(1+s)a_{ij}-s\,b_{ij}=0.
\end{equation}
Now using Eqs. (\ref{newdis}), (\ref{rhopsA}) and (\ref{rhoppsB})
robustness of $\rho$ relative to $\rho^{\prime\prime}_s$ is
defined by
$$
\hspace{-47mm} s=\frac{\parallel\rho-\rho^{\prime}_s\parallel}
{\parallel\rho^{\prime}_s-\rho^{\prime\prime}_s\parallel}
=\sqrt{\left|\frac{Tr(\rho-\rho^{\prime}_s)
(\tilde{\rho}-\tilde{\rho^{\prime}_s})}
{Tr(\rho^{\prime}_s-\rho^{\prime\prime}_s)
(\tilde{\rho^{\prime}_s}-\tilde{\rho^{\prime\prime}_s})}\right|}.
$$
$$
s=\sqrt{\left|\frac{\sum_i(\lambda_i-\lambda^\prime_i)^2+
Tr(AA^\ast)} {\sum_i(\lambda^\prime_i-\lambda^{\prime\prime}_i)^2
+Tr(A-B)(A-B)^\ast}\right|}
$$
\begin{equation}\label{s1AB}
\hspace{-12mm}
=\sqrt{\left|\frac{\sum_i(\lambda_i-\lambda^\prime_i)^2+
Tr(AA^\ast)} {\sum_i(\lambda^\prime_i-\lambda^{\prime\prime}_i)^2+
\frac{1}{s^2}Tr(AA^\ast)}\right|},
\end{equation}
where in the last line Eq. (\ref{eq2AB}) have been used. in order
to show that thus obtained robustness is minimum over all
separable states we need to show that it is local minimum
\cite{vidal}. To do so, it is straightforward to see that the
robustness ${\bf s}$ given in Eq. (\ref{s1AB}) reduces to
\begin{equation}\label{s2AB}
s=\sqrt{\frac{\sum_i(\lambda_i-\lambda^\prime_i)^2}
{\sum_i(\lambda^\prime_i-\lambda^{\prime\prime}_i)^2}},
\end{equation}
which takes minimum values given in Eq. (\ref{rob2}) as the
earlier argument of this section indicates. Since as long as both
the numerator and denominator under the radicals have the same
sign, ${\bf s}$ would be independent of the perturbation of
corresponding separable states. While in cases that numerator and
denominator posses different sign then ${\bf s}$ would be
perturbation dependent. It is trivial to see that the numerator
and denominator should change sign simultaneously, hence ${\bf s}$
would be independent of perturbation. Since both the numerator and
denominator of robustness ${\bf s}$ given in Eq. (\ref{s1AB}) are
continuous function of the matrix elements of matrices $A$ and
$B$, so we expect that ${\bf s}$ is also continuous function of
them. Now suppose that separable states $\rho^\prime_s$ and
$\rho^{\prime\prime}_s$ given in Eqs. (\ref{rhopsA}) and
(\ref{rhoppsB}) are obtained by small perturbation of
corresponding separable states on planes ${\cal S}_1$ and ${\cal
S}^\prime_1$. This means that matrix elements of $A$ and $B$ are
infinitesimal , i.e., $Tr(AA^\ast)$ is infinitesimal. Though
$Tr(AA^\ast)$ can be negative but both numerator and denominator
of Eq. (\ref{s1AB}) remain positive. Therefore for small enough
perturbations the robustness ${\bf s}$ is equal to (\ref{s2AB})
and it is independent of perturbation. Hence if the numerator
changes the sign under the appropriate continuous perturbation
while the denominator remains positive, obviously robustness ${\bf
s}$ would vanishes at perturbation corresponding to the change of
sign of the numerator. Similarly if the denominator changes sign
while the numerator remain positive, the robustness becomes
infinite at the corresponding perturbation. In either case the
robustness ${\bf s}$ as a continuous function of parameters of
perturbation will jump discontinuously to zero or infinite which
is not possible.

In summary as we see the continuous perturbation of separable
states which minimum robustness can not affect the robustness.
This means that the obtained robustness given in Eq. (\ref{rob2})
is local minimum thus according to Ref. \cite{vidal} it is global
minimum.

\section{Evaluation of $K_i$ via the explicit
parameterization of the density matrix} In the previous section we
evaluated robustness for a generic two qubit density matrix. As we
see our approach of evaluation of robustness is based on the
decomposition (\ref{rhox}). As we see thus evaluated robustness is
function of concurrence and also $K_i$, i.e., norm of vectors of
the decomposition. Here in this section we introduce an explicit
parameterization for $\left|x_i\right>$, in order to evaluate
$K_i$. To this aim we define matrix $X$ and $\tilde{X}$ as
\begin{equation}\label{X}
X=\left(\left|x^\prime_1\right>, \left|x^\prime_2\right>,
\left|x^\prime_3\right>, \left|x^\prime_4\right> \right),
\end{equation}
and
\begin{equation}\label{Xtilde}
\tilde{X}=\left(\left|\tilde{x^\prime_1}\right>,
\left|\tilde{x^\prime_2}\right>, \left|\tilde{x^\prime_3}\right>,
\left|\tilde{x^\prime_4}\right> \right),
\end{equation}
respectively. Therefore Eq. (\ref{xxt}) takes the following form
\begin{equation}\label{XtX}
\tilde{X}^\dag X=X^T\sigma_y\otimes\sigma_y X=I.
\end{equation}
Symmetric matrix  $\sigma_y\otimes\sigma_y$ can be diagonalized as
\begin{equation}\label{sigmayy}
\sigma_y\otimes\sigma_y= O^T \eta^2O,
\end{equation}
where $O$ is an orthogonal matrix defined by
\begin{equation}\label{O}
O=\frac{1}{\sqrt{2}}\left(\begin{array}{cccc}
1 & 0 & 0 & 1 \\
0 & 1 & 1 & 0 \\
0 & 1 & -1 & 0 \\
1 & 0 & 0 & -1
\end{array}\right),
\end{equation}
and $\eta$ is the diagonal matrix
\begin{equation}\label{eta}
\eta=\left(\begin{array}{cccc}
i & 0 & 0 & 0 \\
0 & 1 & 0 & 0 \\
0 & 0 & i & 0 \\
0 & 0 & 0 & 1
\end{array}\right).
\end{equation}

Using Eq. (\ref{sigmayy}) we can rewrite Eq. (\ref{XtX}) as
\begin{equation}\label{YtY}
Y^T Y=I,
\end{equation}
where

\begin{equation}\label{YX}
Y=\eta\, O X.
\end{equation}
Equation (\ref{YtY}) shows that Y is a complex 4-dimensional
orthogonal matrix. This means that a given density matrix $\rho$
with corresponding set of positive numbers $\lambda_i$ and
Wootters's basis can transforms under $SO(4,c)$ into a generic
$2\times2$ density matrix with the same set of positive numbers
but with new Wootters's basis. This implies that the space of two
qubit density matrices can be characterize with 12-dimensional (as
real manifold) space of complex orthogonal group $SO(4,c)$
together with four positive numbers $\lambda_i$. Of course the
normalization condition reduces number of parameters to 15.

As far as entanglement is concerned the states $\rho$ and
$\rho^\prime$ are equivalent if they are on the same orbit of the
group of local transformation, that is, if there exist local
unitary transformation $U_1\otimes U_2$ such that
$\rho^\prime=(U_1\otimes U_2)\rho(U_1\otimes U_2)^\dag$, where
$U_1$ and $U_2$ are unitary transformations acting on Hilbert
spaces of particles $A$ and $B$, respectively.

It can be easily seen that under the above mentioned  local
unitary transformations of density matrix $\rho$, the matrix $X$
transforms as
\begin{equation}\label{XprimeX}
X\rightarrow X^\prime=(U_1\otimes U_2)X.
\end{equation}
It is worth to mention that $X^\prime$ also satisfy Eq.
(\ref{XtX}). To show that this is indeed the case, we need to note
that $X^{\prime^T}\sigma_y\otimes\sigma_y X^\prime
=X^T(U_1^T\sigma_y U_1)\otimes(U_2^T\sigma_y U_2)X$. By using
$(\sigma_y)_{ij}=-i\epsilon_{ij}$ we get $(U^T\sigma_y U)_{ij} =
-i\epsilon_{kl}U_{ki}U_{lj}=-i\det(U)\epsilon_{ij}=(\sigma_y)_{ij}$,
where last equality comes from the fact that the unitary matrix
belongs to $SU(2)$. This implies that
\begin{equation}\label{XptXp}
X^{\prime^T}\sigma_y\otimes\sigma_y X^\prime=I.
\end{equation}
By defining $Y^\prime$ as
\begin{equation}\label{YprimeXprime}
Y^\prime=\eta\,OX^\prime,
\end{equation}
one can easily show that $Y^\prime$ is also satisfies
orthogonality condition
\begin{equation}\label{YptYp}
Y^{\prime^T}Y^\prime=I.
\end{equation}
Now by using Eq. (\ref{YprimeXprime}) and inverting Eq.
(\ref{XprimeX}), we can express $Y^\prime$ in terms of $Y$
\begin{equation}\label{YprimeY}
Y^\prime= (\eta\,O)(U_1\otimes U_2)(\eta\,O)^{-1} Y.
\end{equation}
Now by using the fact that $(\eta\,O)\exp({\cal U}_1\otimes {\cal
U}_2)(\eta\,O)^{-1}=\exp((\eta\,O)({\cal U}_1\otimes {\cal
U}_2)(\eta\,O)^{-1})$ and using the explicit form for generators
$({\cal U}_1\otimes {\cal U}_2)$ of local group, one can after
some algebraic calculations see that $(\eta\,O)({\cal U}_1\otimes
{\cal U}_2)(\eta\,O)^{-1}$ is real antisymmetric matrix. This
means that under local unitary transformations matrix $Y$
transforms with $SO(4,r)$ group. So we can parameterize the space
of two qubit density matrices as 6-dimensional coset space
$SO(4,c)/SO(4,r)$ together with 4 positive numbers $\lambda_i$,
which again normailzation condition reduces the number of
parameters to 9.

Below in the rest of this section  we will obtain an explicit
parameterization for a generic two qubit density matrix. First
note that we can decompose coset $SO(4,c)/SO(4,r)$ as
\cite{gilmore}
\begin{equation}\label{cosetdecom}
\frac{SO(4,c)}{SO(4,r)}=\frac{SO(4,c)/SO(4,r)}{SO(2,c)/SO(2,r)\otimes
SO(2,c)/SO(2,r)}\bigotimes
\left(\frac{SO(2,c)}{SO(2,r)}\otimes\frac{SO(2,c)}{SO(2,r)}\right),
\end{equation}
that is, coset representative $Y$ can be decomposed as $Y=Y_1Y_2$.
One can easily show  that coset representative of
$SO(2,c)/SO(2,r)$ has the following form
\begin{equation}
\exp\left(\begin{array}{cc}
0 & i\phi \\
-i\phi & 0 \end{array} \right)= \left(\begin{array}{cc}
\cosh{\phi} & i\sinh{\phi} \\
-i\sinh{\phi} & \cosh{\phi} \end{array} \right).
\end{equation}
Thus $Y_2$ can be written as
\begin{equation}\label{Y2}
Y_2=\left(\begin{array}{c|c}
\begin{array}{cc}
\cosh{\phi_1} & i\sinh{\phi_1} \\
-i\sinh{\phi_1} & \cosh{\phi_1} \end{array} & 0 \\
\hline 0 & \begin{array}{cc}
\cosh{\phi_2} & i\sinh{\phi_2} \\
-i\sinh{\phi_2} & \cosh{\phi_2} \end{array}
\end{array}\right).
\end{equation}
On the other hand $Y_1$ can be evaluated as
\begin{equation}\label{Y1-1}
Y_1=\exp \left( \begin{array}{c|c} 0 & iB \\ \hline -iB^T & 0
\end{array}\right)=
\left( \begin{array}{c|c} \cosh{\sqrt{BB^T}} & iB\frac{\sinh{\sqrt{B^TB}}}{\sqrt{B^TB}} \\
\hline -i\frac{\sinh{\sqrt{B^TB}}}{\sqrt{B^TB}}B^T
 & \cosh{\sqrt{B^TB}}
\end{array}\right)=\left( \begin{array}{c|c} \sqrt{I+CC^T} & iC \\
\hline
 -iC^T & \sqrt{I+C^TC}
\end{array}\right),
\end{equation}
where $B$ is a $2\times2$ matrix and in the last step we used
$C=B\frac{\sinh{\sqrt{B^TB}}}{\sqrt{B^TB}}$. Now using the
singular value decomposition $C=O_1DO^T_2$, Eq. (\ref{Y1-1})
becomes
\begin{equation}\label{Y1-2}
Y_1=\left( \begin{array}{c|c} O_1\sqrt{I+D^2}O^T_1 & iO_1DO^T_2 \\
\hline
 -iO_2DO^T_1 & O_2\sqrt{I+D^2}O^T_2
\end{array}\right),
\end{equation}
where $D$ is a non-negative diagonal matrix. It can be easily seen
that Eq. (\ref{Y1-2}) can be decomposed as
\begin{equation}\label{Y1-3}
Y_1=\left( \begin{array}{c|c} O_1 & 0 \\
\hline
 0 & O_2
\end{array}\right)
\left( \begin{array}{c|c} \sqrt{I+D^2} & iD \\
\hline
 -iD & \sqrt{I+D^2}
\end{array}\right)
\left( \begin{array}{c|c} O^T_1 & 0 \\
\hline
 0 & O^T_2
\end{array}\right).
\end{equation}
By combining Eqs. (\ref{Y2}) and (\ref{Y1-3}) we get
\begin{equation}\label{Y}
Y=\left( \begin{array}{c|c} O_1 & 0 \\
\hline
 0 & O_2
\end{array}\right)
\left( \begin{array}{c|c} \sqrt{I+D^2} & iD \\
\hline
 -iD & \sqrt{I+D^2}
\end{array}\right)
\left( \begin{array}{c|c} O^{\prime}_1 & 0 \\
\hline
 0 & O^{\prime}_2
\end{array}\right).
\end{equation}
Finally using parameterization given in Eq. (\ref{Y2}) we get
$$
Y=\left(\begin{array}{c|c}
\begin{array}{cc}
\cosh{\theta_1} & i\sinh{\theta_1} \\
-i\sinh{\theta_1} & \cosh{\theta_1} \end{array} & 0 \\
\hline 0 & \begin{array}{cc}
\cosh{\theta_2} & i\sinh{\theta_2} \\
-i\sinh{\theta_2} & \cosh{\theta_2} \end{array}
\end{array}\right)
\left( \begin{array}{c|c} \begin{array}{cc}
\cosh{\xi_1} & 0 \\
0 & \cosh{\xi_2} \end{array} &
\begin{array}{cc}
i\sinh{\xi_1} & 0 \\
0 & i\sinh{\xi_2} \end{array} \\
\hline
 \begin{array}{cc}
-i\sinh{\xi_1} & 0 \\
0 & -i\sin{\xi_2} \end{array} &
\begin{array}{cc}
\cosh{\xi_1} & 0 \\
0 & \cosh{\xi_2} \end{array}
\end{array}\right)
$$
\begin{equation}
\left(\begin{array}{c|c}
\begin{array}{cc}
\cosh{\phi_1} & i\sinh{\phi_1} \\
-i\sinh{\phi_1} & \cosh{\phi_1} \end{array} & 0 \\
\hline 0 & \begin{array}{cc}
\cosh{\phi_2} & i\sinh{\phi_2} \\
-i\sinh{\phi_2} & \cosh{\phi_2} \end{array}
\end{array}\right),
\end{equation}
where $\sinh{\xi_i}$ (for $i=1,2$) are diagonal elements of $D$
with the conditions $\xi_i\geq0$.

Using above results and Eq. (\ref{X}) and (\ref{YX}) we can
evaluate the states $\left|x_i\right>$ as

\begin{equation}\label{x1}
{\small\left|x_1\right>=\sqrt{\frac{\lambda_1}{2}} \left(
\begin{array}{c}
-(\sinh{\xi_1}\sinh{\theta_2}\cosh{\phi_1}+\sinh{\xi_2}\cosh{\theta_2}\sinh{\phi_1})
-i(\cosh{\xi_1}\cosh{\theta_1}\cosh{\phi_1}+\cosh{\xi_2}\sinh{\theta_1}\sinh{\phi_1}) \\
-(\sinh{\xi_1}\cosh{\theta_2}\cosh{\phi_1}+\sinh{\xi_2}\sinh{\theta_2}\sinh{\phi_1})
-i(\cosh{\xi_1}\sinh{\theta_1}\cosh{\phi_1}+\cosh{\xi_2}\cosh{\theta_1}\sinh{\phi_1})
\\
(\sinh{\xi_1}\cosh{\theta_2}\cosh{\phi_1}+\sinh{\xi_2}\sinh{\theta_2}\sinh{\phi_1})
-i(\cosh{\xi_1}\sinh{\theta_1}\cosh{\phi_1}+\cosh{\xi_2}\cosh{\theta_1}\sinh{\phi_1})
\\
(\sinh{\xi_1}\sinh{\theta_2}\cosh{\phi_1}+\sinh{\xi_2}\cosh{\theta_2}\sinh{\phi_1})
-i(\cosh{\xi_1}\cosh{\theta_1}\cosh{\phi_1}+\cosh{\xi_2}\sinh{\theta_1}\sinh{\phi_1})
\end{array}
\right)},
\end{equation}
\begin{equation}\label{x2}
{\small\left|x_2\right>=\sqrt{\frac{\lambda_2}{2}} \left(
\begin{array}{c}
(\cosh{\xi_1}\cosh{\theta_1}\sinh{\phi_1}+\cosh{\xi_2}\sinh{\theta_1}\cosh{\phi_1})
-i(\sinh{\xi_1}\sinh{\theta_2}\sinh{\phi_1}+\sinh{\xi_2}\cosh{\theta_2}\cosh{\phi_1})
\\
(\cosh{\xi_1}\sinh{\theta_1}\sinh{\phi_1}+\cosh{\xi_2}\cosh{\theta_1}\cosh{\phi_1})
-i(\sinh{\xi_1}\cosh{\theta_2}\sinh{\phi_1}+\sinh{\xi_2}\sinh{\theta_2}\cosh{\phi_1})
\\
(\cosh{\xi_1}\sinh{\theta_1}\sinh{\phi_1}+\cosh{\xi_2}\cosh{\theta_1}\cosh{\phi_1})
+i(\sinh{\xi_1}\cosh{\theta_2}\sinh{\phi_1}+\sinh{\xi_2}\sinh{\theta_2}\cosh{\phi_1})
\\
(\cosh{\xi_1}\cosh{\theta_1}\sinh{\phi_1}+\cosh{\xi_2}\sinh{\theta_1}\cosh{\phi_1})
+i(\sinh{\xi_1}\sinh{\theta_2}\sinh{\phi_1}+\sinh{\xi_2}\cosh{\theta_2}\cosh{\phi_1})
\end{array}
\right)},
\end{equation}
\begin{equation}\label{x3}
{\small\left|x_3\right>=\sqrt{\frac{\lambda_3}{2}} \left(
\begin{array}{c}
(\sinh{\xi_1}\cosh{\theta_1}\cosh{\phi_2}+\sinh{\xi_2}\sinh{\theta_1}\sinh{\phi_2})
-i(\cosh{\xi_1}\sinh{\theta_2}\cosh{\phi_2}+\cosh{\xi_2}\cosh{\theta_2}\sinh{\phi_2})
\\
(\sinh{\xi_1}\sinh{\theta_1}\cosh{\phi_2}+\sinh{\xi_2}\cosh{\theta_1}\sinh{\phi_2})
-i(\cosh{\xi_1}\cosh{\theta_2}\cosh{\phi_2}+\cosh{\xi_2}\sinh{\theta_2}\sinh{\phi_2})
\\
(\sinh{\xi_1}\sinh{\theta_1}\cosh{\phi_2}+\sinh{\xi_2}\cosh{\theta_1}\sinh{\phi_2})
+i(\cosh{\xi_1}\cosh{\theta_2}\cosh{\phi_2}+\cosh{\xi_2}\sinh{\theta_2}\sinh{\phi_2})
\\
(\sinh{\xi_1}\cosh{\theta_1}\cosh{\phi_2}+\sinh{\xi_2}\sinh{\theta_1}\sinh{\phi_2})
+i(\cosh{\xi_1}\sinh{\theta_2}\cosh{\phi_2}+\cosh{\xi_2}\cosh{\theta_2}\sinh{\phi_2})
\end{array}
\right)},
\end{equation}
\begin{equation}\label{x4}
{\small\left|x_4\right>=\sqrt{\frac{\lambda_4}{2}} \left(
\begin{array}{c}
(\cosh{\xi_1}\sinh{\theta_2}\sinh{\phi_2}+\cosh{\xi_2}\cosh{\theta_2}\cosh{\phi_2})
+i(\sinh{\xi_1}\cosh{\theta_1}\sinh{\phi_2}+\sinh{\xi_2}\sinh{\theta_1}\cosh{\phi_2})
\\
(\cosh{\xi_1}\cosh{\theta_2}\sinh{\phi_2}+\cosh{\xi_2}\sinh{\theta_2}\cosh{\phi_2})
+i(\sinh{\xi_1}\sinh{\theta_1}\sinh{\phi_2}+\sinh{\xi_2}\cosh{\theta_1}\cosh{\phi_2})
\\
-(\cosh{\xi_1}\cosh{\theta_2}\sinh{\phi_2}+\cosh{\xi_2}\sinh{\theta_2}\cosh{\phi_2})
+i(\sinh{\xi_1}\sinh{\theta_1}\sinh{\phi_2}+\sinh{\xi_2}\cosh{\theta_1}\cosh{\phi_2})
\\
-(\cosh{\xi_1}\sinh{\theta_2}\sinh{\phi_2}+\cosh{\xi_2}\cosh{\theta_2}\cosh{\phi_2})
+i(\sinh{\xi_1}\cosh{\theta_1}\sinh{\phi_2}+\sinh{\xi_2}\sinh{\theta_1}\cosh{\phi_2})
\end{array}
\right)}.
\end{equation}
Equations (\ref{x1}) to (\ref{x4}) together with normalization
condition $\sum_{i=1}^4\left<x_i|x_i\right>=1$ give a
parameterization for a generic orbit of two qubit density matrix
up to local unitary group. We are now in position to evaluate
$K_i$. By using Eqs. (\ref{x1}) to (\ref{x4}) after some algebra
we arrive at the following results
$$
K_1=\cosh{2\theta_2}(\sinh^2{\xi_1}\cosh^2{\phi_1}
+\sinh^2{\xi_2}\sinh^2{\phi_1}) \vspace{-5mm}
$$
$$
\vspace{-5mm} \hspace{6mm}
+\cosh{2\theta_1}(\cosh^2{\xi_1}\cosh^2{\phi_1}
+\cosh^2{\xi_2}\sinh^2{\phi_1})
$$
\begin{equation}\label{K1}
\hspace{24mm}
+\sinh{2\phi_1}(\sinh{\xi_1}\sinh{\xi_2}\sinh{2\theta_2}
+\cosh{\xi_1}\cosh{\xi_2}\sinh{2\theta_1}),
\end{equation}
$$
K_2=\cosh{2\theta_2}(\sinh^2{\xi_1}\sinh^2{\phi_1}
+\sinh^2{\xi_2}\cosh^2{\phi_1}) \vspace{-5mm}
$$
$$
\vspace{-5mm} \hspace{6mm}
+\cosh{2\theta_1}(\cosh^2{\xi_1}\sinh^2{\phi_1}
+\cosh^2{\xi_2}\cosh^2{\phi_1})
$$
\begin{equation}\label{K2}
\hspace{24mm}
+\sinh{2\phi_1}(\sinh{\xi_1}\sinh{\xi_2}\sinh{2\theta_2}
+\cosh{\xi_1}\cosh{\xi_2}\sinh{2\theta_1}),
\end{equation}
$$
K_3=\cosh{2\theta_1}(\sinh^2{\xi_1}\cosh^2{\phi_2}
+\sinh^2{\xi_2}\sinh^2{\phi_2}) \vspace{-5mm}
$$
$$
\vspace{-5mm} \hspace{6mm}
+\cosh{2\theta_2}(\cosh^2{\xi_1}\cosh^2{\phi_2}
+\cosh^2{\xi_2}\sinh^2{\phi_2})
$$
\begin{equation}\label{K3}
\hspace{24mm}
+\sinh{2\phi_2}(\sinh{\xi_1}\sinh{\xi_2}\sinh{2\theta_1}
+\cosh{\xi_1}\cosh{\xi_2}\sinh{2\theta_2}),
\end{equation}
$$
K_4=\cosh{2\theta_1}(\sinh^2{\xi_1}\sinh^2{\phi_2}
+\sinh^2{\xi_2}\cosh^2{\phi_2}) \vspace{-5mm}
$$
$$
\vspace{-5mm} \hspace{6mm}
+\cosh{2\theta_2}(\cosh^2{\xi_1}\sinh^2{\phi_2}
+\cosh^2{\xi_2}\cosh^2{\phi_2})
$$
\begin{equation}\label{K4}
\hspace{24mm}
+\sinh{2\phi_2}(\sinh{\xi_1}\sinh{\xi_2}\sinh{2\theta_1}
+\cosh{\xi_1}\cosh{\xi_2}\sinh{2\theta_2}).
\end{equation}

As an example let us consider Bell decomposable states
$\rho=\sum_{i=1}^{4}p_{i}\left|\psi_i\right>\left<\psi_i\right|$,
where $p_i\geq0,\,\sum_i p_i=1$. For these states by choosing
$\theta_1=\theta_2=\xi_1=\xi_2=\phi_1=\phi_2=0$ we get
$\lambda_i=p_i$ and states $\left|x_i\right>$ are given by
\begin{eqnarray}
\left|x_1\right>=-i\sqrt{p_1}\left|\psi_1\right>,\qquad
\left|\psi_1\right>
=\frac{1}{\sqrt{2}}(\left|\uparrow\uparrow\right>+\left|
\downarrow\downarrow\right>),
\\
\left|x_2\right>=\sqrt{p_2}\left|\psi_2\right>,\qquad
\left|\psi_2\right>
=\frac{1}{\sqrt{2}}(\left|\uparrow\downarrow\right>+\left|
\downarrow\uparrow\right>),
\\
\left|x_3\right>=-i\sqrt{p_3}\left|\psi_3\right>,\qquad
\left|\psi_3\right>
=\frac{1}{\sqrt{2}}(\left|\uparrow\downarrow\right>-\left|
\downarrow\uparrow\right>),
\\
\left|x_4\right>=\sqrt{p_4}\left|\psi_4\right>,\qquad
\left|\psi_4\right>
=\frac{1}{\sqrt{2}}(\left|\uparrow\uparrow\right>-\left|
\downarrow\downarrow\right>).
\end{eqnarray}
So it can be easily seen that for these states we get $K_i=1$ for
$i=1,2,3,4$.

\section{Conclusion}
In this work we have obtained robustness of entanglement of a
generic two qubit density matrix.  Our approach to obtain
robustness is based on the decomposition of a generic two qubit
density matrix presented by Wootters.  We have shown that the
robustness of entanglement is proportional to concurrence and in
Bell decomposable density matrices it is equal to the concurrence.
We also present an analytical expression for two separable states
that wipe out all entanglement of these states. We show that
robustness is function of the norm of the vectors of the
decomposition so we present an explicit parameterization for the
decomposition.

\newpage

\vspace{10mm}

{\Large {\bf Figure Captions}}

\vspace{10mm}

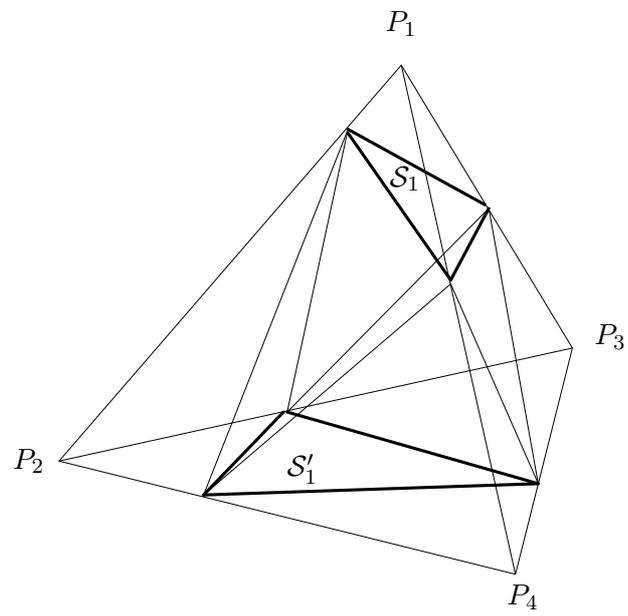
Figure 1: The space of a generic two qubit density matrix is
represented by a tetrahedral. Vertices $P_{i}$ for $i=1,2,3,4$ are
correspond to pure states defined by
$\rho=\lambda_i\left|x^{\prime}_i\right>\left<x^{\prime}_i\right|$.
Irregular octahedral corresponds to separable states. Separable
planes ${\cal S}_1$ and ${\cal S}^\prime_1$ are shown explicitly.

\end{document}